# Massively Parallel Computing of Plasma Turbulence in Tokamaks

**Jeremy Kepner**
Dept. of Astrophysics
Princeton University
jvkepner@astro.princeton.edu

**Scott Parker**
Dept. of Physics
University of Colorado
sparker@buteo.colorado.edu

**Viktor Decyk**
Dept. of Physics
UCLA
vdecyk@pepper.physics.ucla.edu

## Abstract

One of the most important problems in magnetically confined fusion research is the turbulent transport of heat observed in present-day Tokamak experiments. Plasma turbulence is very challenging from a theoretical point of view due to the nonlinearity and high dimensionality of the governing equations. Likewise, experimental measurements in the core region of a Tokamak are limited by the extremely high temperatures $\sim 10^8$ °K. The level of both theoretical and empirical difficulty highlights the important role numerical simulations can potentially play in developing a predictive model for turbulent heat transport. Such a model would dramatically reduce uncertainties in design and may lead to enhanced operating regimes, both of which would reduce the size and expense of a fusion reactor. Simulating this highly non-linear behavior requires solving for the perturbations of the phase space distribution function in five dimensions (via $\delta f$ methods using the gyrokinetic formalism). We use a particle-in-cell approach to solve the equations. The code has been parallelized for a variety of architectures using a 1-D domain decomposition along the toroidal axis, for which the number of particles on each processor remains approximately constant—minimizing load imbalances and interprocessor communication—making this problem ideally suited to massively parallel architectures. We present the performance of the program for different numbers of processors and problem sizes.

## Geometry

The essential geometry of a Tokamak is that of a torus defined by a major radius R and a minor radius a (see **Figure 1**). The ions within the plasma move rapidly around the torus, gyrating tightly along the magnetic field lines, like rings on a wire. The radius of gyration $\rho$ is set by the velocity, mass and charge of the particle. The essential scale of the system is set by $a/\rho$. Typically R ~ 270 cm, a ~ 85 cm, and $\rho$ ~ 0.15 cm.

The simplest arrangement of field lines is obtained by wrapping current carrying wires tightly around the minor axis of the torus creating straight magnetic field lines aligned with the torus. Unfortunately, the magnetic field exerts a greater force on the inside of the torus, which causes the ions in the plasma to drift across the field lines. This problem can be partially alleviated by twisting the field lines into a helical shape so that the drifts approximately cancel. A byproduct of this twisting is that the trajectories of particles are now far more complex and are susceptible to a wide range of instabilities, which tend to grow along the toroidal modes of the Tokamak. **Figure 2** shows a simplified example of a typical toroidal mode. This could be, for example, the linear growth phase of an ion density perturbation. Generally, the isosurfaces of a perturbation follow the helical shape of the magnetic field lines.

## Implementation

Particle-in-cell codes have been used in the plasma physics community for several decades. The essence of the computational problem involves solving for the trajectories of particles with mass m and charge q using equations analogous to

$$d\mathbf{v}_i(t)/dt = (q/m)\, \mathbf{E}(\mathbf{x}_i(t)) , \qquad d\mathbf{x}_i(t)/dt = \mathbf{v}_i(t) ,$$

$$\nabla \cdot \mathbf{E} = 4\pi\, q\, n(\mathbf{x}) , \qquad n(\mathbf{x}) = \sum_i S(\mathbf{x} - \mathbf{x}_i) ,$$

where the subscript i refers to the ith particle and the electric field at the particle's position is $\mathbf{E}(\mathbf{x}_i(t))$ is found by interpolating from fields previously calculated on a grid. The interpolation is a gather operation which involves indirect addressing and a substantial part of the computation time is spent here. The fields created by the particles are found from Poisson's equation. This equation is solved on a grid, using Fourier transform methods. Typically, the time for the field solver is not large. The source term $n(\mathbf{x})$ is calculated from the particle position by an inverse interpolation where S is a particle shape function. This is a scatter operation which also involves indirect addressing and consumes a substantial part of the computation time.

## Results

When the simulations start with a very small initial perturbation, there are generally two identifiable phases of the run. First is the linear phase, where modes (i.e., standing waves) grow exponentially. Linear modes can also be found using lower-dimensional, time-independent eigenvalue techniques and are fairly well understood theoretically. The second phase is the turbulent stationary state, during which the growth of the dominant linear modes saturates and the system settles down to a statistical steady state. The ion density in the linear mode is shown on surfaces throughout the Tokamak in **Figure 3**.

In our implementation, a 1D domain decomposition along the toroidal axis is used, which is significantly easier to program than a full 3D decomposition. In addition, along this axis the number of particles per processor remains relatively constant, which minimizes load imbalance. The most recent efforts have been in porting the code to the Cray T3D using Fortran77 and the PVM message passing library. Production runs on the T3D show a performance of approximately 14.4 MFlops per processor (10% of the theoretical peak of 150 MFlops, which is typical of these types applications). To test the scalability of the code, 12 runs were timed with different numbers of processors and problem sizes. To first order, the problem size is given by the total number of particles. The total amount of computer resource consumed is the time per step multiplied by the number of processors. In a perfectly scalable code, the resource consumption should be proportional to the problem size, which is shown by the straight line in **Figure 4**.

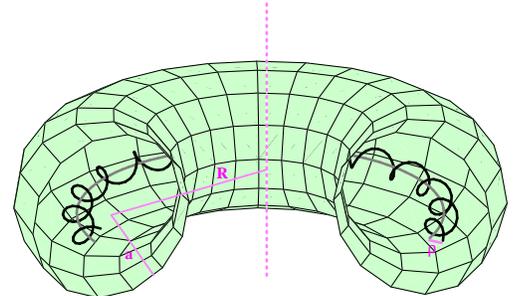

**Figure 1.** Schematic drawing of a Tokamak with major radius R and minor radius a, and the gyrating path of a particle (black line) along a magnetic field line (gray line) with a gyration radius $\rho$.

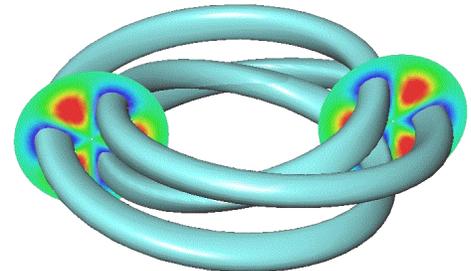

**Figure 2.** Example of a m=3, n=2 mode in a torus. m is the poloidal mode number and n is the toroidal mode number. The isosurface shows the geometry of a typical growing standing wave within the plasma, it also traces the magnetic field lines.

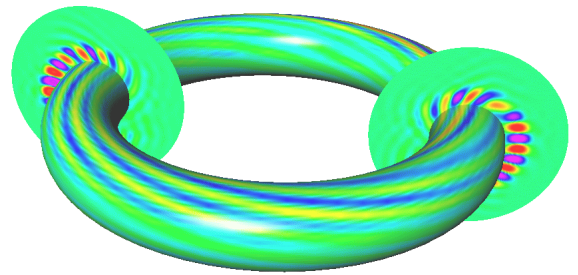

**Figure 3.** Ion density in the linear phase of the simulation showing the ion-temperature-gradient instability. The mode is very elongated in the direction following the magnetic field lines.

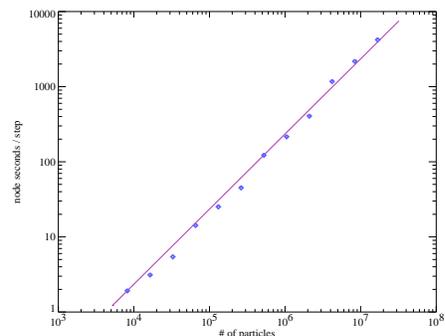

**Figure 4.** Total processor time used vs. problem size. The straight line indicates the expected time from scaling up the smallest simulation.